\documentclass[10pt,letterpaper]{article}
\usepackage{opex3}

\begin{document}

\title{Wideband, Efficient Optical Serrodyne Frequency Shifting with a Phase Modulator and a Nonlinear Transmission Line}

\author{Rachel Houtz$^{2}$, Cheong Chan$^{1}$ and Holger M\"{u}ller}

\address{Department of Physics, 261, Birge Hall, University of California at Berkeley \\ Berkeley, CA, 94720 USA}

\email{$^{1}$c2@berkeley.edu, $^{2}$rhoutz@berkeley.edu}

\begin{abstract}
We report shifting of the frequency of an 850 nm laser with an
instantaneous bandwidth of (350-1650)\, MHz and an efficiency
between 35\% (minimum) to 80\% (best at frequencies around 600 and
1500\, MHz) by phase modulation with a sawtooth waveform
(``serrodyne frequency shifting"). We use a fiber-coupled
traveling wave electro-optical modulator driven by a nonlinear
transmission line.
\end{abstract}

\ocis{(060.5060) Phase modulation (060.5625) Radio frequency
photonics (130.0250) Optoelectronics (130.3730) Lithium niobate}

\section{Introduction}
Shifting the frequency of a laser is important for many
applications in, e.g., atomic, molecular and optical physics,
optical measurement, laser frequency and phase stabilization
\cite{Hall}, optical gyroscopes \cite{Laskoskie}, coherent optical
communication systems \cite{Voges}, or spectroscopy. Tunable
lasers can often replace such frequency shifters especially when a
large frequency shift is desired \cite{Fujiwara}. However, this
technique may require phase-locking, be costly, and have a slow
response time. Acousto-optic modulators (AOM) \cite{Hobbs} are
commercially available for frequencies between $\sim
30-2500$\,MHz, and achieve high efficiency especially for models
up to 200 MHz. However, their tuning range is limited to a
relatively narrow ($\pm 10\%$) band around the center frequency of
each individual model, even if the deflection angle change is
compensated for in a double-pass configuration. A good example for
what can be achieved with such double-pass AOMs is described by
Donley \textit{et al.} where the peak efficiency is 75\% and the
tuning bandwidth is 68 MHz full width at half maximum power for
light coupled into a single-mode fiber \cite{Donley}. On the other
hand, electro-optical phase modulators (EOMs) can be used to
transfer power into sidebands that are symmetric around the
frequency of the laser. The frequency shift of the sidebands is in
principle variable, but the maximum power transferred to any
output frequency is limited to about 33\% with sine wave
modulation \cite{Hobbs}.

Optical serrodyne frequency shifting theoretically allows one to
achieve a wideband frequency shift with high efficiency
\cite{Johnson}. The idea is to use an EOM for applying phase
modulation with a sawtooth waveform. Driving the EOM with a
perfect sawtooth waveform in theory shifts the frequency of the
optical signal with complete suppression of undesired sidebands
\cite{Johnson} and 100 \% efficiency. Previously shifts with
electronically generated sawtooth waveforms have been reported in
the MHz range \cite{Laskoskie,Johnson,Wong}. A larger shift of up
to 1.28 GHz has been reported with a generation of the sawtooth
waveform that involved a photonic arbitrary waveform generator
\cite{Chou}. However, this not only requires the use of a
femtosecond laser, but also limits the agility of the frequency
changes, since the sawtooth period depends on the repetition rate
of the laser pulses. Optical serrodyne frequency shifting has been
utilized to achieve heterodyne self-mixing in a
distributed-feedback fiber laser for use in a coherent detection
system \cite{Laroche}. Serrodyne frequency shifters based on
multistage phase modulation have been demonstrated
\cite{Hisatake}.

Here, we describe an entirely electronic method of generating the
sawtooth waveform, utilizing a nonlinear transmission line (NLTL).
It has the benefit of being much simpler and cheaper to set up
than the above photonic method while achieving even larger shifts.
We report frequency shifting of an 850 nm laser by (350-1650)\,
MHz using a fiber-coupled travelling wave electro-optical
modulator driven by a NLTL. We experimentally demonstrate an
efficiency of 35\% (minimum) to 80\% (best at select frequencies
around 600 and 1500\, MHz). Within these frequency ranges, the
magnitude of the frequency shift can be instantaneously tuned
because it is given by an rf sine wave generator.

\section{Description}
Figure 1 shows the experimental setup. It can be broken down into
two main parts: First, the serrodyne shifter, which consists of
the electronics and the EOM and second, an interferometer that we
use to characterize the frequency shifting by a beat note
measurement.

\begin{figure}[htbp]
\centering\includegraphics[width=14cm]{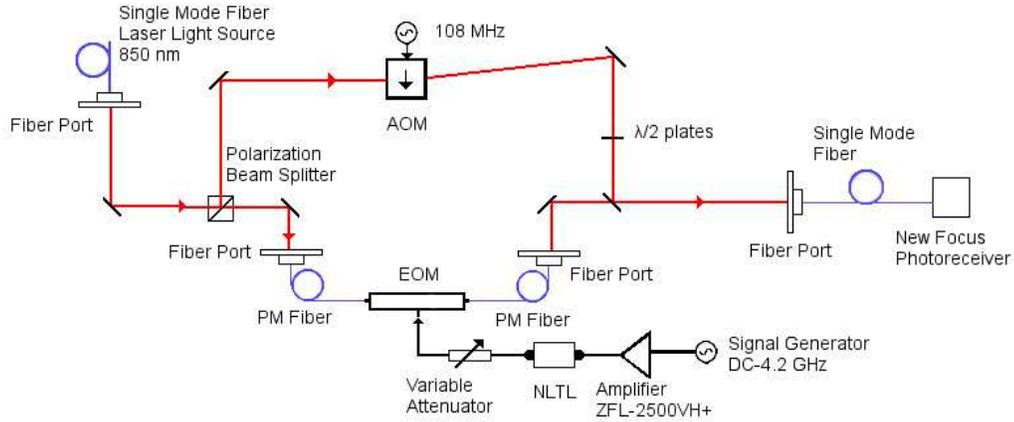}
\caption{Diagram of Experimental setup}
\end{figure}

Our serrodyne shifter consists of the fiber-coupled EOM and the
sawtooth generator based on the NLTL and associated drive
electronics. The EOM is a lithium niobate modulator
(PM-0K5-10-PFU-PFU-850-UL by EOSpace) with pigtailed
polarization-maintaining fibers connected to both the input and
output of the crystal. Our particular modulator is specified for a
0-12 GHz bandwidth with less than 2 dB variation in modulation
efficiency. The half-wave voltage is specified as 2.3 V at 1 GHz.
The specified optical insertion loss is 2.4 dB, corresponding to
58\% overall optical transmission from input to output; we
typically achieve 50\%. The EOM's rf input is driven by the signal
from the output of a monolithic gallium arsenide nonlinear
transmission line (model 7103 and 7113 ``comb generators'' by
Picosecond Pulse Labs). In the basic setup (see Figure 1), the
NLTL is driven with a sine wave created by a HP 8665A Signal
Generator that is amplified to $\sim 20$ dBm using a Mini-circuits
ZFL-2500VH+ amplifier.

The NLTL is basically a ladder of inductors and semiconductor
diodes which function as voltage-variable capacitors
\cite{afshari,bond}. When a sine wave propagates through the NLTL,
the higher voltage parts of the signal propagate faster than the
lower voltage parts. This is because the capacitance of the diodes
is reduced by the applied voltage which, in turn, speeds up the
signal propagation through the NLTL. This dispersion distorts the
sine waveform input into an approximate sawtooth waveform. The
polarity of the diodes in our NLTLs is such that positive-going
ramps are obtained. Obtaining negative ones would require a NLTL
with the opposite polarity.

We use optical heterodyne detection to measure the frequency shift
induced by the electro-optic modulator. We use laser light from a
grating-stabilized diode laser at a wavelength of 850 nm that
arrives at the experiment via a single mode fiber optic cable.
Approximately half of this light is split off by a polarization
beam splitter and coupled into the serrodyne shifter, with its
polarization matched to the EOM.

The other half of the laser power is frequency shifted by
$\omega_{\rm AOM}=-2\pi\times 108$\,MHz with an acousto-optical
modulator and overlapped with the serrodyne shifter's output on a
beam splitter, where the two beams interfere. A New Focus 12 GHz
amplified photoreceiver detects the heterodyned signal. The output
signal of the photoreceiver is connected to a HP 8591A spectrum
analyzer with an input range of DC-1.8 GHz. In our particular
setup, the frequency shift of the EOM is positive, $+\omega_{\rm
EOM}$, so the observed beat frequency by the spectrum analyzer
will be $\omega_{\rm AOM}+\omega_{\rm EOM}$.

\section{Results}
We measure the efficiency of the frequency translation as the
ratio of the power in the desired sideband to the power of the
carrier without the phase modulation. Note that this definition
does not include fiber coupling losses. For some applications,
fiber coupling is required anyway; in this case, there is very
little additional loss since the EOM with fibers is almost as
efficient as a bare fiber (50-80\%, depending on the quality of
the input beam). For optical serrodyne frequency shifting, the
efficiency will depend on how well the amplitude (peak value) of
the RF input matches the half wave voltage of the phase modulator,
and any distortions in the sawtooth \cite{Udd}.

We first characterize the efficiency obtained with the 7103 NLTL
driven by at a constant amplitude sine wave signal. Figure 2
depicts the efficiency as a function of the frequency shift with
sine wave input signals for 3 constant amplitudes. Using a drive
of -3 dBm, we achieved an efficiency of 60\% and a bandwidth of
200-1500 MHz full width at half maximum for the shifted power.

\begin{figure}[h]
\centering\includegraphics[width=12cm, height=6cm]{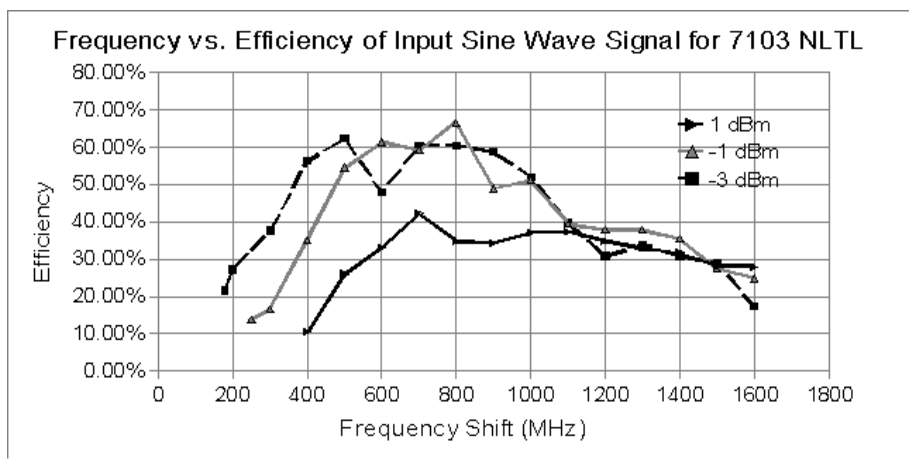}
\caption{Efficiency measured using the 7103 NLTL versus frequency
for 3 different constant amplitude sine wave signals from the
signal generator}
\end{figure}

The efficiency data shown in gray in Figure 3 were obtained using
the model 7103 NLTL. For each frequency, we optimized the
efficiency by varying the amplitude of the sine wave input to the
amplifier (between -10 to +13 dBm). For frequency shifts in the
range 350 MHz to 1.65 GHz, we obtain more than 35\% efficiency.
The maximum efficiency (at 1250 MHz) was 61\%.

The data shown in black in Figure 3 were taken using the 7113
model NLTL. We again optimized the amplitude of the input sine
wave for each frequency. We also found that by varying the power
supply voltage between 9 and 15 V for the Mini-circuits
ZFL-2500VH+ RF amplifier, an improved efficiency could be
obtained. This improvement of performance is supposedly due to the
distortion of the sine wave by the amplifier which helps create a
better sawtooth voltage at the NLTL output. We achieved a 1.5 GHz
frequency shift with 80\%. Relative to constant-amplitude drive
(Fig. 2), these data show either higher peak efficiency ($>$70\%
in bands around 600, 800, 1000 and 1500 MHz) or greater uniformity
($>$40\% for 350-1650 MHz).

\begin{figure}[h]
\centering\includegraphics[width=12cm, height=6cm]{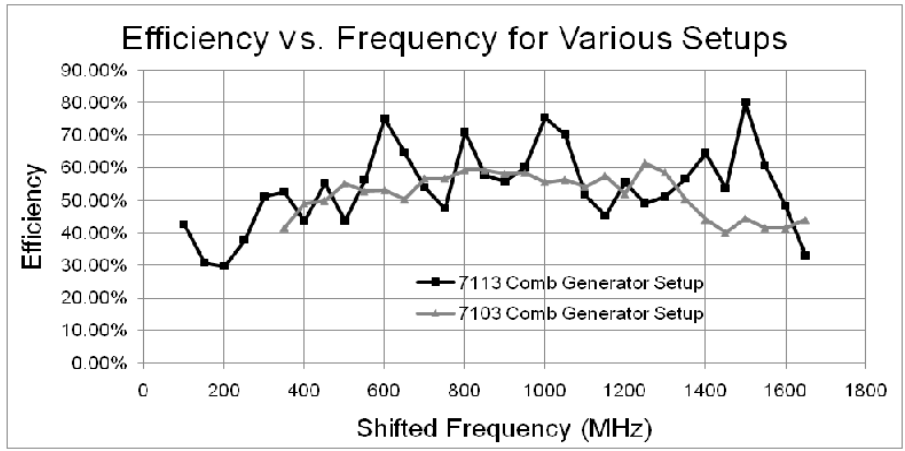}
\caption{Performance of the two NLTLs when the input waveform is
optimized for each frequency as described in the text}
\end{figure}

For shifts below 1.25 GHz, a further improvement in efficiency
could be achieved by adding the first harmonic of the drive
frequency to the sine wave with adjustable amplitude and phase.
This results in a rough approximation of a sawtooth delivered to
the input of the NLTL, which then generates a higher quality
sawtooth at its output, see figure 4. The harmonic is generated
using a frequency doubler (ANZAC DI-4). To adjust the amplitude
and phase, we use a variable delay line and a variable attenuator.
The fundamental frequency and the harmonic are combined in a
Mini-circuits power combiner (ZFSC-2-2). The combined signal is
amplified with a Mini-circuits ZFL-1000VH and drives the model
7103 NLTL which is connected directly to the EOM RF port.

\begin{figure}[htbp]
\centering\includegraphics[width=6cm]{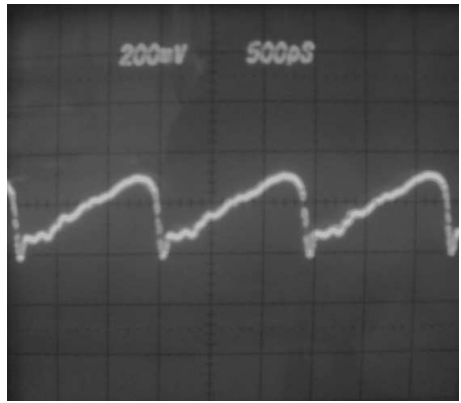}
\caption{Waveform after the 7103 NLTL as measured by a Tektronix
7904 oscilloscope equipped with an S-2 sampling plug-in (75\,ps
specified risetime) with a 20 dB attenuator. Input to the setup
was 700\,MHz sine wave. The obtained efficiency of the resulting
frequency shift was 68\%.}
\end{figure}

Although in this configuration, the amplifier and the power
combiner are used far beyond their specified frequency range, we
achieved 80\% efficiency at 600 MHz and better than 50\%
efficiency in the range of 450 MHz to 1.25 GHz. At higher
frequencies, our amplifier cannot pass the harmonic because its
frequency is far beyond the specifications of the amplifier. For
each data point, we optimized the amplitude of the input sine wave
(12 to 20 dBm), the input power to the amplifier (7 to 15 V), the
length of the linear delay line, and the attenuation on the
variable attenuator (0 to 7 dB), see figure 5.

\begin{figure}[htbp]
\centering\includegraphics[width=12cm, height=6cm]{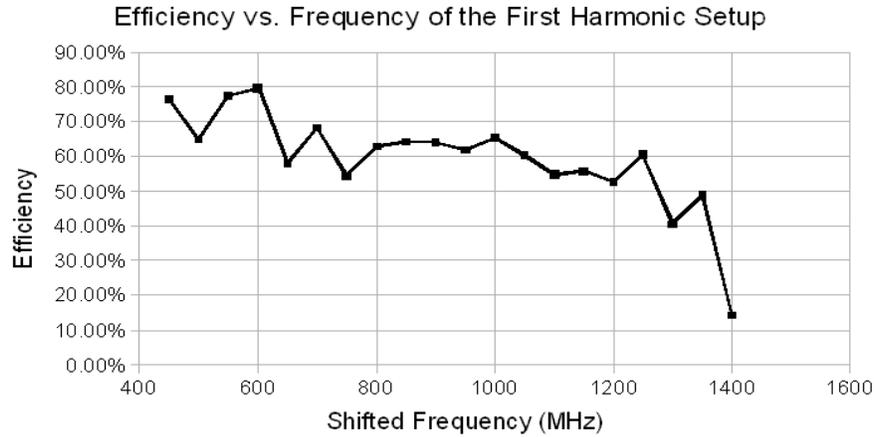}
\caption{Graph of the performance of the 7103 comb generator with
first harmonic mixed into the input}
\end{figure}

\paragraph{Distribution of the power among sidebands}
The power not shifted into the desired sideband is distributed
between the original input frequency and other (undesired)
sidebands. In one particular measurement, this distribution (in
dB) was measured as shown in Tab. \ref{dist}. The setup used for
taking this data had a small frequency shift of 8.2\,MHz, to allow
for accurate data taking out to the highest sidebands. However,
the data should be representative of the distribution of the power
in serrodyne shifting when the efficiency is about 75\%.

\begin{table}
\centering \caption{\label{dist} Power in dB referred to the total
power for various sidebands. The -1.26 dB for the 1$^{\rm st}$
sideband indicates a 75\% shifting efficiency.}
\begin{tabular}{ccccccccccccc}\hline\hline
Sideband \# & 2 & 1 & 0 & -1 & -2 & -3 & -4 & -5 & -6 & -7 &-8 &-9 \\
Power [dB] & -20& -1.26& -17&
-15&-14&-15&-15&-17&-17&-17&-18&-18\\ \hline \hline
\end{tabular}
\end{table}

\section{Conclusion}
In summary, we have presented a wideband serrodyne frequency
shifter based on a lithium niobate phase modulator and a nonlinear
transmission line. We achieved frequency shifting of a 850 nm
laser by with efficiencies ranging from 35\% (minimum) to 80\%
over 1.5 GHz bandwidth. The current limitation on electronic
sawtooth production is still the limiting factor in this class of
phase modulators. This type of frequency shifter is ideally
suitable to maintain constant detuning for large atom
interferometry experiments and has many other potential
applications.

\paragraph{Note Added:} A similar implementation of
serrodyne frequency shifting has been reported in Ref.
\cite{mark}.

\section{Acknowledgements}
We thank the Undergraduate Research Apprentice Program (URAP) of
UC Berkeley.


\begin{thebibliography}{99}

\bibitem{Hall} J. L. Hall and T. W. H\"{a}nsch,
``External dye-laser frequency stabilizer," Opt. Lett. \textbf{9}, 502-504 (1984).

\bibitem{Laskoskie} C. Laskoskie, H. Hung, T. El-Wailly, and C. L. Chang, ``Ti-LiNbO
waveguide serrodyne modulator with ultrahigh sideband suppression
for fiber optic gyroscopes," J. Lightwave Technol. \textbf{4},
600–606 (1989).

\bibitem{Voges} E. Voges and K. Petermann, \textit{Optische Kommunikationstechnik} (Springer, Berlin, 2002).

\bibitem{Fujiwara} N. Fujiwara et al.,
``140-nm Quasi-Continuous Fast Sweep Using SSG-DBR Lasers," IEEE
Photon. Technol. Lett. \textbf{20}, 1015 (2008).

\bibitem{Hobbs} P. C. D. Hobbs, \textit{Building Electro-optical Systems} (Wiley, New York, 2000).

\bibitem{Donley} E. A. Donley, T. P. Heavner, F. Levi, M. O. Tataw, and S. R. Jefferts,
``Double-pass acousto-optic modulator system," Rev. Sci. Instrum. \textbf{76}, 063112 (2005).

\bibitem{Johnson} L. M. Johnson and C. H. Cox, ``Serrodyne optical frequency translation
with high sideband suppression," J. Lightwave Technol. \textbf{6},
109–112 (1988).

\bibitem{Wong} K. K. Wong and R. M. De La Rue,
``Electro-optic-waveguide Frequency Translator in LiNbO3
Fabricated by Proton Exchange," Opt. Lett. \textbf{7}, 546-548
(1982).

\bibitem{Chou} I. Y. Poberezhskiy, B. Bortnik, J. Chou, B. Jalali, and H. R. Fetterman,
``Serrodyne Frequency Translation of Continuous Optical Signals
Using Ultrawide-band Electrical Sawtooth Waveform," IEEE J.
Quantum Electron. \textbf{41}, 1533-1539 (2005).

\bibitem{Laroche} Mathieu Laroche, Célia Bartolacci, Guillaume Lesueur, Herv\'e Gilles, and Sylvain Girard,
``Serrodyne optical frequency shifting for heterodyne self-mixing
in a distributed-feedback fiber laser," Opt. Lett. \textbf{33},
2746-2748 (2008).

\bibitem{Hisatake} S. Hisatake and T. Kobayashi,
``Electro-optic transparent frequency conversion of a continuous light
wave based on multistage phase modulation," Opt. Lett. \textbf{31}, 498-500 (2006).

\bibitem{Afshari} E. Afshari and A. Hajimiri, ``A non-linear transmission line for
pulse shaping in silicon," IEEE J. Solid-State Circuits,
\textbf{40}, 744–752 (2005).

\bibitem{Bond} Bradley N. Bond and Luca Daniel,
``A Piecewise-Linear Moment-Matching Approach to Parameterized
Model-Order Reduction for Highly Nonlinear Systems," IEEE
Transactions on Computer-Aided Design Of Integrated Circuits And
Systems, \textbf{26}, 2116 (2007).

\bibitem{Udd} E. Udd,
\textit{Fiber Optic Sensors - An Introduction for Engineers and Scientists} (John Wiley and Sons, New York, 2006).

\bibitem{mark} D. M. S. Johnson, J. M. Hogan, S. Chiow, M. A. Kasevich,
``Broadband Optical Serrodyne Frequency Shifting",
arXiv:0909.1834.

\end{thebibliography}
\end{document}